\documentclass[aps,prb,twocolumn,superscriptaddress,floatfix,longbibliography]{revtex4-2}

\usepackage{amsmath,amssymb} 
\usepackage{bm} 
\usepackage{graphicx} 
\usepackage{comment} 
\usepackage{textcomp} 

\usepackage[version=4]{mhchem} 
\usepackage{balance} 
\usepackage{enumitem}
\setlist{noitemsep,leftmargin=*,topsep=0pt,parsep=0pt}
\usepackage[dvipsnames]{xcolor}
\usepackage[normalem]{ulem}

\definecolor{lightgray}{gray}{0.6}
\definecolor{medgray}{gray}{0.4}

\usepackage{hyperref}
\hypersetup{
  colorlinks=true,
  linkcolor=blue,
  citecolor=blue,
  urlcolor=blue,
}

\newif\ifptitle
\newif\ifpnumber
\newcounter{para}

\usepackage{float}

\usepackage{xspace}
\newcommand{\Td}{\ensuremath{\mathrm{T_d}}\xspace}  

\newcommand{\mytitle}{Stacking-Induced Tuning of a Flat-Band Surface State in \ce{NbIrTe4}}

\begin{document}

\title{\mytitle}

\author{Jiabin Yu}
\affiliation{Quantum Matter Institute, University of British Columbia, Vancouver, BC, V6T 1Z4, Canada}
\affiliation{Department of Physics and Astronomy, University of British Columbia, Vancouver, BC, V6T 1Z1, Canada}

\author{Mohamed Oudah}
\affiliation{Quantum Matter Institute, University of British Columbia, Vancouver, BC, V6T 1Z4, Canada}

\author{Giang Nguyen}
\affiliation{Quantum Matter Institute, University of British Columbia, Vancouver, BC, V6T 1Z4, Canada}

\author{Hsianghsi Kung}
\affiliation{Quantum Matter Institute, University of British Columbia, Vancouver, BC, V6T 1Z4, Canada}

\author{V. King}
\affiliation{Quantum Matter Institute, University of British Columbia, Vancouver, BC, V6T 1Z4, Canada}
\affiliation{Department of Chemistry, University of British Columbia, Vancouver, British Columbia V6T 1Z1, Canada}

\author{Rysa Greenwood}
\affiliation{Quantum Matter Institute, University of British Columbia, Vancouver, BC, V6T 1Z4, Canada}
\affiliation{Department of Physics and Astronomy, University of British Columbia, Vancouver, BC, V6T 1Z1, Canada}

\author{Adrian F. Rumson}
\affiliation{Department of Chemistry, Dalhousie University, Halifax, NS, B3H 4R2, Canada}

\author{Jisun Kim}
\affiliation{Quantum Matter Institute, University of British Columbia, Vancouver, BC, V6T 1Z4, Canada}

\author{Matteo Michiardi}
\affiliation{Quantum Matter Institute, University of British Columbia, Vancouver, BC, V6T 1Z4, Canada}
\affiliation{Department of Physics and Astronomy, University of British Columbia, Vancouver, BC, V6T 1Z1, Canada}

\author{Giorgio Levy}
\affiliation{Quantum Matter Institute, University of British Columbia, Vancouver, BC, V6T 1Z4, Canada}
\affiliation{Department of Physics and Astronomy, University of British Columbia, Vancouver, BC, V6T 1Z1, Canada}

\author{Andrea Damascelli}
\affiliation{Quantum Matter Institute, University of British Columbia, Vancouver, BC, V6T 1Z4, Canada}
\affiliation{Department of Physics and Astronomy, University of British Columbia, Vancouver, BC, V6T 1Z1, Canada}

\author{E. R. Johnson}
\affiliation{Department of Chemistry, Dalhousie University, Halifax, NS, B3H 4R2, Canada}

\author{Meigan Aronson}
\affiliation{Quantum Matter Institute, University of British Columbia, Vancouver, BC, V6T 1Z4, Canada}
\affiliation{Department of Physics and Astronomy, University of British Columbia, Vancouver, BC, V6T 1Z1, Canada}

\author{Lukas Muechler}
\affiliation{Department of Chemistry, Penn State University, University Park, PA 16802, U.S.A}

\author{Sarah A. Burke}
\email[]{saburke@phas.ubc.ca}
\affiliation{Quantum Matter Institute, University of British Columbia, Vancouver, BC, V6T 1Z4, Canada}
\affiliation{Department of Physics and Astronomy, University of British Columbia, Vancouver, BC, V6T 1Z1, Canada}
\affiliation{Department of Chemistry, University of British Columbia, Vancouver, British Columbia V6T 1Z1, Canada}

\date{\today}

\begin{abstract}

Ternary layered transition metal dichalcogenides (TMDCs) offer high tunability and access to correlated and topological states. However, like the broader family of layered TMDCs, their properties are sensitive to stacking order and interlayer coupling. Here, we investigate the electronic structure and surface states of \ce{NbIrTe4}, a candidate Weyl semimetal in bulk, and predicted to be a quantum spin Hall insulator in monolayer form. Using scanning tunnelling microscopy (STM), angle-resolved photoemission spectroscopy (ARPES), and density-functional theory (DFT), we attribute the strong resonance in tunnelling spectroscopy located approximately 8\,meV below the Fermi level to a surface-derived flat band which shows sensitivity to the local stacking order at the surface. In regions where the top layer is laterally shifted, tunnelling spectroscopy instead shows a gap-like feature with intensity near the Fermi level shifted in energy. Combining DFT and quasiparticle interference (QPI) measurements, we find this dramatic change in the spectroscopy in the vicinity of the quasi-1D flat-band surface state becomes dispersive due to a change in the spatial distribution of the density of states in the top two layers. These findings establish \ce{NbIrTe4} as a platform where surface band dispersion can be effectively tuned through local structural manipulation.

\end{abstract}

\maketitle

 \section{\label{sec:Intro}Introduction}

Transition metal dichalcogenides (TMDCs) have long provided a versatile platform for exploring symmetry-broken and correlated phases~\cite{wilson1969transition, yoffe1973layer, wilson1975charge}. Members of this material class frequently host electronic instabilities  and nontrivial band topology, giving rise to a diverse phase diagram that includes charge-density waves (CDWs)~\cite{wilson1975charge,soumyanarayanan2013quantum}, superconductivity~\cite{klemm2015pristine}, and topological  states such as type-II Weyl semimetals~\cite{soluyanov2015type,li2017evidence}.
Due to the weak interlayer van der Waals (vdW) coupling, a defining characteristic of TMDCs is the tunability of their electronic structure via stacking order and layer thickness~\cite{mak2010atomically,splendiani2010emerging,wang2012electronics,manzeli20172d}.
Within the bulk phases, different polymorphs consisting of different stacking orders exhibit different electronic properties~\cite{wilson1969transition,yang2017structural}, while fabrication of monolayer, bilayer, and twisted stacking have demonstrated the incredible power of modifying interlayer coupling and stacking symmetry~\cite{chen2020strong,andrei2021marvels,mak2022semiconductor,barre2024engineering}.

Moving beyond binary TMDCs, ternary compounds of layered transition metal chalcogenides further expand the phase space for correlated and topological phenomena.
The integration of an additional transition metal species introduces extra orbital degrees of freedom and reduced lattice symmetry, resulting in a complex and tunable electronic structure~\cite{xing2020surface,shipunov2021layered,lee2024spin}.
In the prototypical type-II Weyl semimetal \ce{TaIrTe4}, angle-resolved photoemission spectroscopy (ARPES) and transport measurements have revealed a nontrivial topology characterized by Weyl points and Fermi-arc surface states~\cite{koepernik2016tairte,belopolski2017signatures,dong2019observation}.

The sister compound $\text{NbIrTe}_4$ is isostructural to $\text{TaIrTe}_4$, adopting the noncentrosymmetric orthorhombic space group $Pmn2_1$~\cite{li2017ternary,liu2017van}.
This distorted octahedral coordination leads to structural anisotropy and broken inversion symmetry, hosting a variety of exotic electronic phases. Notably, bulk \ce{NbIrTe4} exhibits non-saturating magnetoresistance at ambient pressure~\cite{zhou2019nonsaturating,schonemann2019bulk,huang2019magneto}, and pressure-induced superconductivity has also been reported~\cite{mu2021pressure,long2021observation}, highlighting the sensitivity of its electronic properties to subtle lattice modifications and structural tuning.

\ce{NbIrTe4} is predicted to transition from a Weyl semimetal in bulk, to a quantum spin Hall insulator (QSHI) in the monolayer limit, a phase transition driven by changes in the vdW interlayer coupling~\cite{liu2017van,liu2025prediction}, similar to \Td-WTe$_2$~\cite{qian2014quantum,jiang2015signature,zheng2016quantum}.
Prior studies have indicated that accessing such exotic states may not strictly require physical exfoliation as local variations in stacking order and the associated symmetry breaking can fundamentally alter the band topology, such as in the case of graphite, where altering the Bernal stacking decouples the surface layer and recovers monolayer-like Dirac-dispersion~\cite{li2007observation,li2009scanning,luican2011quantized}, while a recent THz STM study showed evidence of a topological phase transition coupled to shear of the topmost atomic layer of bulk \ce{WTe2}~\cite{jelic2025terahertz}. However, given the range of outcomes apparent from changes in stacking order of TMDCs, the effects of modifying interlayer coupling require careful examination.

In this work, we present a combined experimental and theoretical study of the surface electronic structure of \ce{NbIrTe4}.
Using scanning tunneling microscopy (STM), we identify a prominent surface state resonance just below the Fermi level.
Complementary ARPES measurements reveal that this feature originates from a quasi-1D flat band along the $\Gamma$--$Y$ direction, giving rise to a sharp enhancement in the density of states.
Notably, we find that this sharp spectral signature vanishes in regions where the top surface layer is laterally shifted, demonstrating the critical sensitivity of the electronic structure to the local stacking order.
Our analysis shows that the suppression arises because the lateral shift modifies the interlayer coupling, causing the flat band to evolve into a highly dispersive state, consistent with the emergence of quasi two-dimensional electronic behavior.

\begin{figure}
    \includegraphics[clip=true,width=1\columnwidth]{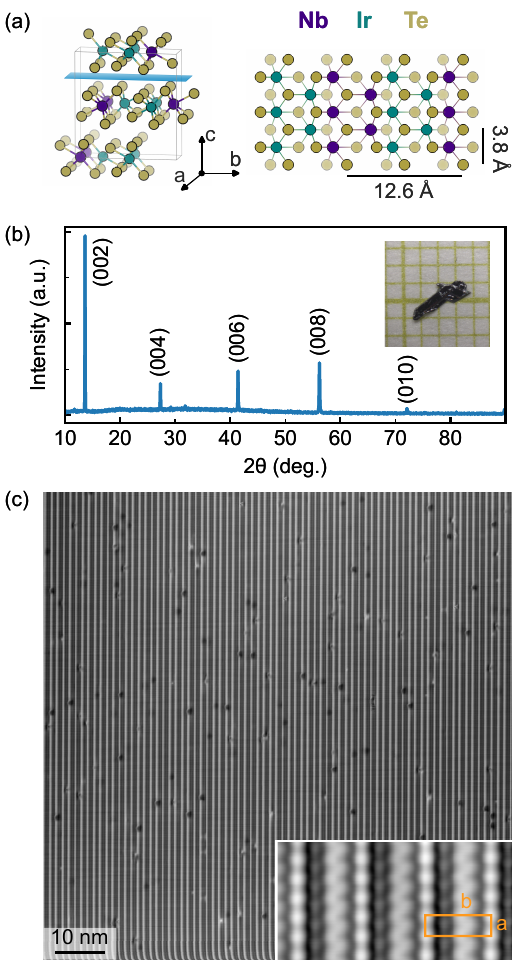}
    \caption{
        Structural and topographic characterization of \ce{NbIrTe4}.
        (a) Perspective (left) and top (right) view of the unit cell; the cyan plane indicates the natural cleavage plane.
        (b) Single-crystal XRD pattern corresponding to the expected structure.
        Inset: Optical image of a typical needle-like crystal. Grid size: $1\,\text{mm}$.
        (c) Large-scale STM topograph ($V_{\mathrm{B}} = -0.2\,\mathrm{V}$, $I_{\mathrm{t}} = 200\,\mathrm{pA}$, $\Delta z=1.23~\text{\AA}$) showing flat terraces with parallel 1D chains. Inset: atomic-resolution image ($V_{\mathrm{B}} = 0.1\,\mathrm{V}$, $I_{\mathrm{t}} = 200\,\mathrm{pA}$, $4.2 \times 2.2\,\mathrm{nm}^2$, $\Delta z = 4.47~\text{\AA}$) with the unit cell marked in orange.
             }
    \label{fig:overall_sample_info}
\end{figure}

\section{\label{sec:methods}Methods}

Single crystals of \ce{NbIrTe4} were grown using a \ce{Te} self-flux method. High-purity \ce{Nb}, \ce{Ir}, and \ce{Te}  were loaded into a Canfield crucible set  and sealed under vacuum in a quartz tube. The sealed tubes were heated to $1323\,\mathrm{K}$ over 5~hours, held at this temperature for 24~hours, and then slowly cooled to $973\,\mathrm{K}$ over 80~hours. The samples were characterized by x-ray diffraction (XRD), which revealed a well-defined single-crystalline phase consistent with the orthorhombic lattice of \ce{NbIrTe4}.

After growth, the crystals were transferred into an ultra-high vacuum (UHV) system and cleaved in situ before STM measurements. The base pressure of the chamber was better than $5 \times 10^{-11}\,\mathrm{mbar}$. \ce{PtIr} tips were calibrated on a Ag(111) surface prior to use.

ARPES measurements were performed at the UBC-Moore Center for Ultrafast Quantum Matter using a Scienta DA-30L hemispherical analyzer and the 21.2 eV photons (He I) generated by a monochromatized Scienta Omicron VUV5k helium discharge lamp. The samples were cleaved and measured below $10$~K in a vacuum better than $8 \times 10^{-11}\,\mathrm{mbar}$. The total energy and angular resolutions were better than $4$~meV and $0.2^{\circ}$, respectively.

Electronic structure calculations were performed by density-functional theory (DFT) using the VASP package~\cite{VASP} with the default PBE-PAW-5.4
pseudopotentials for Ir and Te, while the s-valence potential was chosen for Nb. The experimental geometries were taken from the ICSD. For the self-consistent calculations, the reducible Brillouin zone (BZ) was sampled by a $9\times3\times3$ k-mesh. A Wannier interpolation of 208 bands was performed by projecting onto an atomic-orbital basis centered at the atomic positions, consisting of valence Nb-$d$, Ir-$d$ as well as Te-$s$ and Te-$p$ orbitals. Spin-orbit coupling was included for all calculations. For the slab calculations, a $1 \times 1 \times 8$ supercell was constructed with a vacuum of 34~\text{\AA} in the $c$-direction.

Structural relaxations to investigate changes in interlayer spacing were performed using two functionals with lightdenser basis sets in FHI-aims version 250822
\cite{blum2009ab,abbott2025roadmap}: B86bPBE\cite{becke1986large} and
PBE\cite{perdew1996generalized}, both paired with the exchange-hole dipole moment (XDM) dispersion
correction\cite{price2023xdm} with Z
damping\cite{becke2024remarkably,bryenton2026consistent}. The lattice
vectors were fixed during the relaxation, allowing only the atomic
postions to change until no force on any atom exceeded 0.01
eV/\text{\AA}. A k-grid of $10 \times 6 \times 3$ was used for the bulk geometries, with
a k-grid of $10 \times 6 \times 1$ for the slabs with the shifted layers. Scalar
relativistic effects were described with the zeroth-order regular
approximation\cite{van1994relativistic}.  

\section{\label{sec:results}Results}

\subsection{Crystal structure and topography}

The layered structure of \ce{NbIrTe4} consists of distorted octahedral planes stacked along the $c$-axis (Fig.~\ref{fig:overall_sample_info}a). The lattice constants $a = 3.8\,\text{\AA}$ and $b = 12.6\,\text{\AA}$ are consistent with prior studies~\cite{chen2019unusual, sun2025unveiling}. The natural cleavage plane, located between the weakly bonded \ce{Te-Te} layers, is depicted by the cyan plane in Fig.~\ref{fig:overall_sample_info}a.
A typical high-quality, ribbon-like metallic single crystal is shown in the inset of Fig.~\ref{fig:overall_sample_info}b.
The crystal structure was confirmed by single-crystal XRD after growth (Fig.~\ref{fig:overall_sample_info}b), which demonstrates the high crystallinity of the sample, matching the structural characteristics with reported literature~\cite{zhou2019nonsaturating,huang2019magneto}. Following \textit{in-situ} cleavage, a large, flat surface area was obtained, maintaining an atomically clean environment for topographic and spectroscopic mapping.

Fig.~\ref{fig:overall_sample_info}c displays a representative STM topograph of the vacuum-cleaved \ce{NbIrTe4} surface, revealing wide, atomically flat terraces.
The observed one-dimensional modulations are consistent with the alternating Ir and Nb chains within the $ab$-plane (Fig.~\ref{fig:overall_sample_info}a), which are further resolved in the high-resolution topographic image inset of Fig.~\ref{fig:overall_sample_info}c. A low density of point defects indicates a pristine surface, enabling high-resolution spectroscopic and quasiparticle interference (QPI) measurements.

\subsection{Flat band near the Fermi level}

\begin{figure}
    \includegraphics[clip=true,width=1\columnwidth]{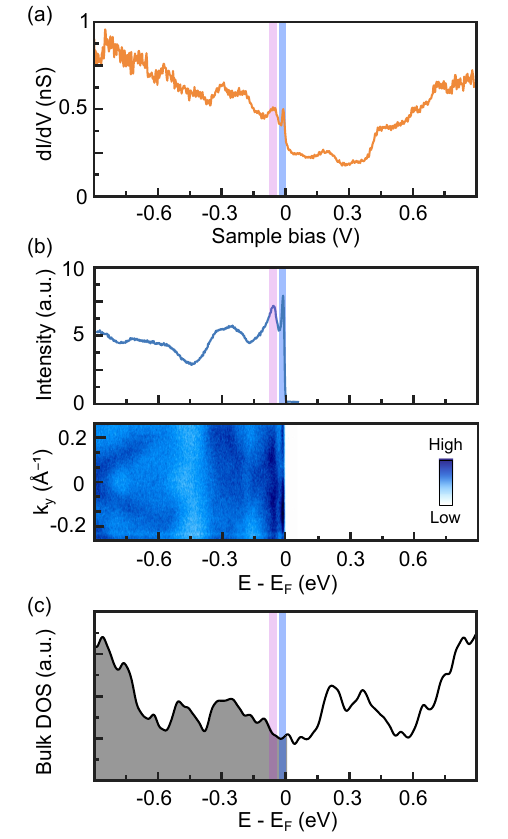}
    \caption{
        Spectroscopic evidence for surface states in \ce{NbIrTe4}.
        (a) Spatially averaged $dI/dV$ spectrum from a pristine terrace, revealing a sharp resonance at $-8\,\mathrm{meV}$(light blue) and a broader feature near $-50\,\mathrm{meV}$(light pink).
        (b) ARPES characterization along the $\Gamma$--$Y$ direction.
        Bottom: Intensity map depicting two nearly dispersionless (flat) bands. 
        Top: Integrated EDC displaying prominent peaks corresponding to the surface bands.
        (c) Calculated bulk DOS from DFT (occupied states shaded gray).
        The absence of distinct peaks near $E_\text{F}$ in the bulk calculation, in contrast to (a) and (b), supports the surface origin of the observed resonances.
     }
    \label{fig:bulkDOS_dIdV_ARPES}
\end{figure}

To probe the local electronic structure, we performed scanning tunneling spectroscopy (STS) on the atomically flat terraces.
The spatially averaged $dI/dV$ spectrum reveals a prominent, sharp peak with a full width at half maximum (FWHM) of $\sim 10\,\text{meV}$ near the Fermi level.
This intense resonance,  centered around $-8\,\text{meV}$ (Fig.~\ref{fig:bulkDOS_dIdV_ARPES}a), has been previously attributed to a van Hove singularity \cite{wang2022observation}.
A secondary, broader feature is resolved around $-50$ meV, consistent with the more dispersing surface state reported by Ekahana \textit{et al.}~\cite{ekahana2020topo}. Temperature-dependent STS measurements (Fig.~S1), further show persistence of the near-Fermi peak up to $40\,\text{K}$ ruling out low-temperature correlated phases as may arise from the Kondo effect.

To identify the origin of these spectral features, we performed ARPES measurements to map the momentum-resolved electronic structure.
The photoemission intensity map along the $\Gamma$--$Y$ direction (Fig.~\ref{fig:bulkDOS_dIdV_ARPES}b) reveals two bands located at binding energies of $-8$ and $-50\,\text{meV}$. These two bands show little dispersion in this direction but do disperse along $\Gamma$--$X$, characteristic of quasi-1D electronic structure. 
The corresponding energy distribution curve (EDC)  displays distinct peaks at these energies, in excellent agreement with the tunneling resonances observed in the STS spectrum (Fig.~\ref{fig:bulkDOS_dIdV_ARPES}a). Comparing now to the bulk density of states (DOS) calculated via DFT, we see a metallic continuum lacking any distinct peaks near the Fermi level (Fig.~\ref{fig:bulkDOS_dIdV_ARPES}c).
The absence of these features in the bulk DOS suggests they are inherently surface-derived bands.

\subsection{Stacking-dependent electronic structure}

Beyond the pristine surface regions discussed above, we also identify domains characterized by lateral stacking displacements.
Such structural variations originating from growth dynamics~\cite{katzke2004phase} or cleavage-induced strain are commonly observed in layered vdW, as previously reported in graphite~\cite{li2009scanning} and TMDCs such as \ce{WSe2}~\cite{pedramrazi2019manipulating}, \ce{NbSe2}~\cite{soumyanarayanan2013quantum}, and \ce{WTe2}~\cite{wang2019formation}, as well as induced by locally enhanced THz fields~\cite{jelic2025terahertz}.
These local deviations offer a direct pathway to probe how surface electronic states respond to specific stacking configurations.

\begin{figure}
    \includegraphics[clip=true,width=1.0\columnwidth]{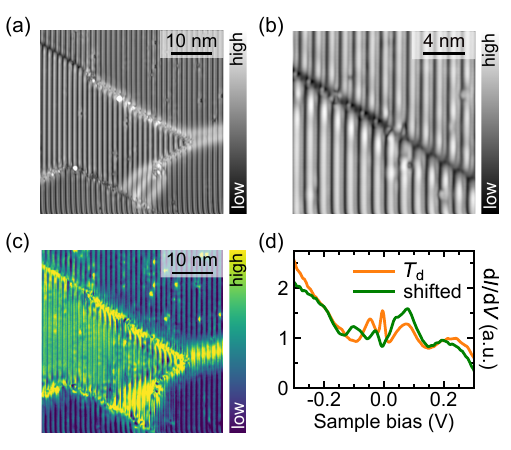}
    \caption{
        Electronic structure of laterally displaced surface regions in \ce{NbIrTe4}.
            (a) Large-scale STM topograph ($V_{\mathrm{B}} = -0.1\,\mathrm{V}$, $I_{\mathrm{t}} = 200\,\mathrm{pA}$, $\Delta z=1.9~\text{\AA}$) revealing a domain boundary where the surface lattice exhibits a half-unit-cell lateral offset relative to the surrounding area.
            (b) Zoomed-in topograph of the same region ($V_{\mathrm{B}} = -0.1\,\mathrm{V}$, $I_{\mathrm{t}} = 300\,\mathrm{pA}$, $\Delta z=1.6~\text{\AA}$) highlighting the lattice displacement across the boundary.
            (c) Corresponding $dI/dV$ map of the same area ($V_{\mathrm{B}} = -0.1\,\mathrm{V}$, $I_{\mathrm{t}} = 200\,\mathrm{pA}$), displaying distinct contrast between the shifted and pristine \Td regions.
            (d) Spatially averaged $dI/dV$ spectra extracted from the shifted (green) and pristine \Td (orange) regions. The sharp resonance near $E_\text{F}$ characteristic of the \Td phase is suppressed into a gap-like feature in the shifted region.
        }
    \label{fig:SurfaceShifted}
\end{figure}

A representative topographic image (Fig.~\ref{fig:SurfaceShifted}a) and a zoomed-in view across the interface (Fig.~\ref{fig:SurfaceShifted}b) reveal a sharp domain boundary with a half-unit-cell lateral offset in the \textit{b}-direction of the surface layer. This structural shift profoundly modifies the local electronic structure. As shown in the $dI/dV$ map at $V = -0.1\text{ V}$ (Fig.~\ref{fig:SurfaceShifted}c), the local density of states (LDOS) exhibits strong spatial contrast across the boundary. Spatially averaged spectroscopy (Fig.~\ref{fig:SurfaceShifted}d) shows this contrast results from significant changes in the low-energy spectral features: the sharp resonance characteristic of the pristine \Td stacking phase (orange curve) is suppressed with a gap-like appearance near $E_\text{F}$ within the shifted domains (green curve). These features are observed in multiple similar half-unit cell shifted regions, and are thus expected to be a result of the change in stacking order.

\begin{figure*}
    \centering
    \includegraphics[width=\textwidth]{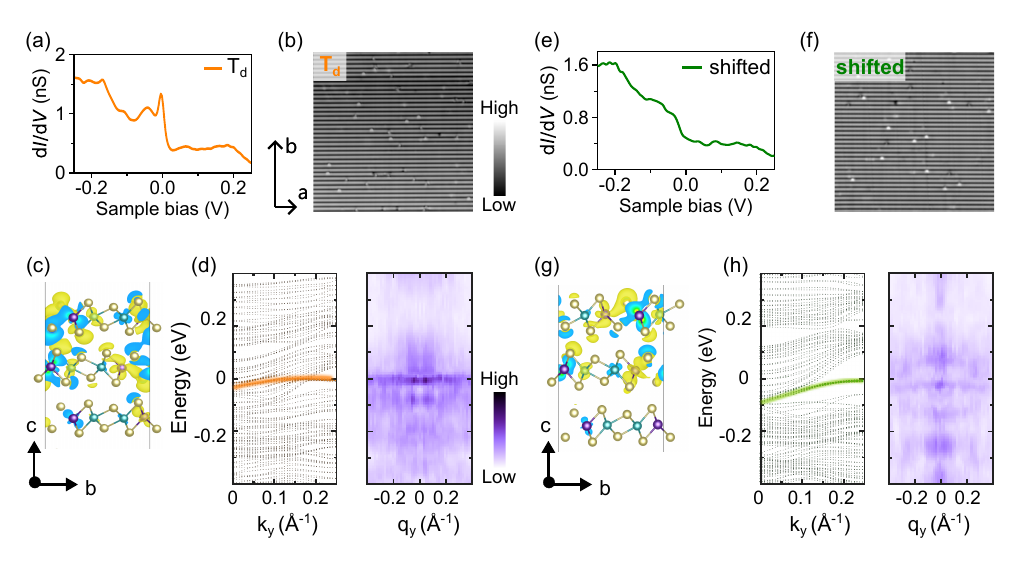}
     \caption{ Stacking-dependent electronic structure modulation in \ce{NbIrTe4} probed by STS, DFT, and QPI.
            Panels (a) to (d) correspond to the pristine \Td stacking and panels (e) to (h) to the shifted stacking.
            (a, e) Spatially averaged $dI/dV$ spectra ($N = 110$ each), extracted from spectroscopic grids acquired on the STM topographs in (b) and (f). The shifted domain in (f) was measured at a separate location. Both topographs are uniform and were acquired with $I_{\mathrm{t}} = 500\,\mathrm{pA}$ and $V_{\mathrm{B}} = -400\,\mathrm{mV}$.
            (c, g) Calculated charge density distributions in the $bc$ plane.
            (d, h) DFT band structures along $k_y$ and QPI line cuts along $q_y$. The highlighted bands reproduce the measured dispersions and show a clear change in band dispersion between the two stackings.
           }
    \label{fig:SurfaceStacking}
  \end{figure*}

To uncover the microscopic origin of this change, DFT calculations were performed for both stacking configurations (Fig.~\ref{fig:SurfaceStacking}d,h).
These electronic structure calculations were performed on the unrelaxed geometry based on XRD of the bulk, however, structural optimization including dispersion corrections (as described in the methods) show minimal geometry changes for the shifted stack, and notably only a 0.05\,\AA~outward relaxation of the interlayer spacing compared to bulk, indicating a minimal change in the interlayer coupling.
For the electronic structure calculations, both stacking configurations produce surface states observed near $E_\text{F}$; however, the shifted slab configuration shows much higher dispersion in the $\Gamma-Y$ direction. This particularly broadens and shifts the previously prominent sharp peak near $E_F$. Examination of the calculated charge-density distribution (Fig.~\ref{fig:SurfaceStacking}c,g) corresponding to this surface state shows that it spans the top two layers, and that the shift results in a re-organization of the electron density: instead of the high density regions being stacked vertically and remaining in well-isolated 1-dimensional rows, the change in stacking order creates a 2-dimensional character that bridges the two layers. This more 2-dimensional character straightforwardly accounts for the increased band dispersion.
These slab calculations reveal that the lateral displacement alters the character of the local interlayer coupling, with minimal influence on coupling strength, driving a surface-band reconstruction that disrupts their quasi-1D character.

To experimentally confirm that this geometry-induced change in the band dispersion is responsible for the changes in spectra, rather than more exotic possibilities such as QSHI phase, QPI measurements were performed in regions with and without sharp surface state feature (Fig.~\ref{fig:SurfaceStacking}(a-d) and (g-h) respectively) and compared to the dispersion calculated by DFT. A mixed-domain region containing both phases is shown in Fig.~S2.
The sharp \Td resonance and the gap-like suppression in the shifted phase remain highly robust across different areas and with different microscopic tip terminations.

The momentum-space character of these spectral modifications is revealed by comparing the DFT-calculated band dispersions along $k_y$ with the experimental QPI line cuts along $q_y$ (Fig.~\ref{fig:SurfaceStacking}d,h).
For the \Td domain, the QPI line cut (Fig.~\ref{fig:SurfaceStacking}d, right) clearly resolves a nearly flat band feature near the Fermi level ($E_\text{F}$). This lack of dispersion is responsible for the high-DOS resonance observed in the tunneling spectra and aligns closely with the calculated band structure, featuring a surface state with weak dispersion in the $\Gamma-Y$ direction (Fig.~\ref{fig:SurfaceStacking}d, left). This flat band also agrees well with the ARPES measurements shown in Fig.~\ref{fig:bulkDOS_dIdV_ARPES}b. In contrast, the QPI signature in the shifted region reveals that this state exhibits pronounced dispersion (Fig.~\ref{fig:SurfaceStacking}h, right), corresponding well with the increased dispersion of the DFT-calculated surface state for the shifted stacking configuration (Fig.~\ref{fig:SurfaceStacking}h, left).

This transition from a flat to a dispersive band driven by surface stacking order provides a direct momentum-space verification for the origin of the suppression of the high-DOS peak. 
The lateral displacement of the top layer of \ce{NbIrTe4} effectively reshapes the surface band structure through a fundamental change in symmetry and dimensionality of the electronic coupling between the topmost layers. 
This shift in dimensionality from a nearly flat 1D-like band to a dispersive state explains the quenching of the sharp resonance at $E_\text{F}$. Meanwhile, emergence of a gap-like structure does not clearly point to a QSHI state or other changes in topology, but instead arises from slight shifts in the band edges. 
Here, lateral stacking order emerges as a powerful degree of freedom for tuning the dimensionality of the surface electronic structure of \ce{NbIrTe4}.

\section{Conclusion}
\label{sec:conclusion}

Here, we have identified a surface state that produces a sharp resonance in tunnelling spectroscopy just below the Fermi level in \ce{NbIrTe4}. This feature arises from a quasi-1D surface-derived flat band as confirmed by ARPES, DFT, and QPI. This quasi-1D surface state  exhibits dramatic sensitivity to stacking order and symmetry. In contrast to other 2D materials, where stacking faults may lead to decoupling of the top layer~\cite{li2007observation,li2009scanning,luican2011quantized}, here we do not attribute the change in electronic structure to monolayer-like behaviour such as the emergence of the predicted QSHI state.
Instead, we attribute the change in the local density of states to a qualitative difference in the interlayer coupling resulting in a more 2D character, leading to increased dispersion and broadening of the density of states near the Fermi energy. Such lateral stacking displacements act as a switch, creating a more 2D electron distirbution, dispersing the flat band surface state associated with the \Td phase. This stacking-controlled reconstruction of a surface-derived band is complementary to recent demonstrations in \ce{WTe2}, where THz-induced top-layer shear is coupled to a change in band topology~\cite{jelic2025terahertz} suggesting potential avenues for control of electronic structure and topology.

\section*{ACKNOWLEDGMENTS}

This research was undertaken thanks in part to funding from the Max Planck–UBC–UTokyo Centre for Quantum Materials and the Canada First Research Excellence Fund, Quantum Materials and Future Technologies. This project is also funded by the Natural Sciences and Engineering Research Council of Canada (NSERC), Canada Foundation for Innovation(CFI); the Department of National Defence (DND); the British Columbia Knowledge Development Fund (BCKDF); the Gordon and Betty Moore Foundation’s EPiQS Initiative, Grant GBMF4779 to A.D.; the Canada Research Chairs Program (A.D.); and the CIFAR Quantum Materials Program (A.D.).
We thank the Atlantic Computational Excellence Network (ACEnet) for computational resources.


\appendix

\bibliography{reference}

\end{document}